\documentclass[twocolumn]{aastex63} % twocolumn,linenumbers, manuscript

\received{August 31, 2020}
\revised{October 14, 2020}
\accepted{October 19, 2020}
\submitjournal{ApJL}

\shorttitle{Coronal extrapolation from synchronic data with AI-generated farside}
\shortauthors{Jeong et al.}

\graphicspath{{./}{figures/}}

\begin{document}

\title{Solar Coronal Magnetic Field Extrapolation from Synchronic Data with AI-generated Farside}

\author[0000-0003-4616-947X]{HYUN-JIN JEONG}
\affiliation{School of Space Research, Kyung Hee University, Yongin, 17104, Republic of Korea}

\correspondingauthor{Yong-Jae Moon}
\email{moonyj@khu.ac.kr}

\author[0000-0001-6216-6944]{Yong-Jae Moon}
\affiliation{School of Space Research, Kyung Hee University, Yongin, 17104, Republic of Korea}
\affiliation{Department of Astronomy and Space Science, College of Applied Science, Kyung Hee University, Yongin, 17104, Republic of Korea}

\author[0000-0003-0969-286X]{Eunsu Park}
\affiliation{Department of Astronomy and Space Science, College of Applied Science, Kyung Hee University, Yongin, 17104, Republic of Korea}

\author[0000-0002-9300-8073]{Harim Lee}
\affiliation{Department of Astronomy and Space Science, College of Applied Science, Kyung Hee University, Yongin, 17104, Republic of Korea}

\begin{abstract}
Solar magnetic fields play a key role in understanding the nature of the coronal phenomena.
Global coronal magnetic fields are usually extrapolated from photospheric fields, for which farside data is taken frontside, about two weeks earlier.
For the first time we have constructed the extrapolations of global magnetic fields using frontside and artificial intelligence (AI)-generated farside magnetic fields at a near-real time basis.
We generate the farside magnetograms from three channel farside observations of Solar Terrestrial Relations Observatory (STEREO) Ahead (A) and Behind (B) by our deep learning model trained with frontside Solar Dynamics Observatory extreme ultraviolet images and magnetograms.
For frontside testing data sets, we demonstrate that the generated magnetic field distributions are consistent with the real ones; not only active regions (ARs), but also quiet regions of the Sun.
We make global magnetic field synchronic maps in which conventional farside data are replaced by farside ones generated by our model.
The synchronic maps show much better not only the appearance of ARs but also the disappearance of others on the solar surface than before.
We use these synchronized magnetic data to extrapolate the global coronal fields using Potential Field Source Surface (PFSS) model.
We show that our results are much more consistent with coronal observations than those of the conventional method in view of solar active regions and coronal holes.
We present several positive prospects of our new methodology for the study of solar corona, heliosphere, and space weather.
\end{abstract}


%%%%%%%%%%%%%%%%%%%%%%%%%%%%%%%%%%%%%%%%%%%%%%%%%%%%%%%%%%%%%%%%%%%%%%%%%%%%%%%%%

\section{Introduction} \label{sec:intro}

Solar magnetic fields dominate the structure and dynamics of the corona and inner heliosphere \citep{Jess2016}.
The magnetic field is an energy source of solar flares and their accompanying coronal mass ejections \citep{Amari2018,Inoue2018}.
Photospheric magnetic fields are routinely measured, but direct measurements of coronal magnetic fields are very difficult and rare.
Thus, several models of extrapolation or magnetohydrodynamic (MHD) simulation have been developed to derive coronal magnetic fields from photospheric magnetograms \citep{Mikic2018,Nandy2018}.
A synoptic map of solar magnetic fields has been widely used for the input boundary condition of the coronal models.
The map constructed by merging together frontside magnetograms, which are daily updated near-central-meridian data, over 27-day solar rotation period \citep{Bertello2014}.
For the farside of the map, data assimilation techniques with magnetic surface flux transport models and helioseismic farside detections were applied \citep{DeVore1984,Schrijver2003}.
However, those approaches still have limitations to predict real-time farside magnetic fields, especially for rapid changes in magnetic fields by flux emergence or disappearance near the limb.

Deep learning, a subset of machine learning in artificial intelligence (AI) and also known as deep neural networks, has been developed to find the best mathematical manipulation to turn the input into the output, whether it be a linear or nonlinear relationship.
And it has made many advances to solve space weather problems like solar flare forecast \citep{Huang2018,Park2018}, coronal hole detection \citep{Illarionov2018}, etc..
The \textit{Pix2Pix} \citep{Isola2017} model, which is based on the conditional Generative Adversarial Networks  \citep[cGAN;][]{Mirza2014}, is a novel deep learning method excellent for image translation tasks that have been well demonstrated in solar astronomy and space weather \citep{Park2019,Park2020,Ji2020}.
\cite{Kim2019} (hereafter KPL19) proposed an approach to generate solar farside magnetograms from the Solar TErrestrial RElations Observatory  \citep[STEREO;][]{Kaiser2008}/Extreme UltraViolet Imager \citep[EUVI;][]{Howard2008} 304 {\AA} images.
The orbit of STEREO Ahead (A) and Behind (B) is at a distance of about 1 AU, and drift away from the Earth at a rate of about $22 ^{ \circ}$ per year in opposite directions.
They applied the \textit{Pix2Pix} and trained and evaluated the model with pairs of frontside Solar Dynamics Observatory \citep[SDO;][]{Pesnell2012}/Atmospheric Imaging Assembly \citep[AIA;][]{Lemen2012} 304 {\AA} images and SDO/Helioseismic and Magnetic Imager \citep[HMI;][]{Schou2012} line-of-sight (LOS) magnetograms before the generation of farside data.
However, their results were limited to the maximum magnetic field strength of $\pm 100$ Gauss, and showed low correlations in solar quiet regions.
It was noted that the \textit{Pix2Pix} has an issue with generating high-resolution data, and the lack of details and realistic features in the high-resolution results \citep{Chen2017}.
\cite{Wang2018} proposed an improved model to solve the issues with a novel adversarial loss and multi-scale architectures of the networks, and they named the model "\textit{Pix2PixHD}".
\cite{Shin2020} tried to generate magnetograms from Ca II K images, and their results show that the \textit{Pix2PixHD} model is useful to generate ones with a large dynamic range ($\pm1400$ Gauss).

In this Letter, we make an upgraded model with $\pm3000$ Gauss dynamic range based on the \textit{Pix2PixHD} and use multi-channel images for input.
We describe the detailed structure of the model in Section \ref{sec:pix2pixHD}.
We train the model with frontside SDO extreme ultraviolet (EUV) passband images and magnetograms, and then generate farside magnetograms from the corresponding images of STEREO A and B by the model in Section \ref{sec:farside}. 
Next we generate a synchronic map of photospheric field, which replaces the conventional synoptic map at the farside with the AI-generated ones in Section \ref{sec:synchronic}.
Hereafter, this is called the HMI \& AI synchronic map.
Finally, we extrapolate coronal magnetic fields from the synchronic map using a Potential Field Source Surface (PFSS) model.
Then we compare them with EUV observations as well as those from the conventional method in Section \ref{sec:extrapolation}.
We present several prospects of our results in Section \ref{sec:summary}. 

%%%%%%%%%%%%%%%%%%%%%%%%%%%%%%%%%%%%%%%%%%%%%%%%%%%%%%%%%%%%%%%%%%%%%%%%%%%%%%%%

\section{Data} \label{sec:Data}

We use pairs of full-disk SDO/AIA three passband images and SDO/HMI LOS 720s magnetograms to train our deep learning model.
The three EUV passbands are 304, 193, and, 171 {\AA} corresponding to the chromosphere, corona, and upper transition region, respectively \citep{Lemen2012}.
The data pairs have a cadence of 12 hr (at 00 UT and 12 UT each day) from 2011 January 1 to 2017 December 31.
We use 4231 pairs of the multi-channel EUV images for input and magnetograms for output of our model.
We construct a training set with 10 months and an evaluation set with two months, and both are selected for each year without any duplication between the two sets.
In order to train and evaluate various inclination conditions, which cause different distributions of southern/northern magnetic field strength for each month \citep{Pastor2015}, the months are selected randomly.
We take 3412 pairs for the training data set and 819 pairs for the evaluation data set.

To generate farside magnetograms, we use STEREO/EUVI A \& B (304, 195, and 171 {\AA}) passband images, which have similar response characteristics to the SDO/AIA images \citep{Downs2012}.
These SDO and STEREO passbands are often used together for the global solar studies \citep{Su2012,Cairns2018}.
The farside EUV images are selected from the closest times (within one hour) to the synoptic data, which is a conventional boundary condition for the coronal magnetic field extrapolation.

The following data pre-processing is applied to the EUV data and magnetograms for the effective training and generating.
We make Level 1.5 images with the standard SolarSoftWare (SSW) packages \citep{Freeland1998} of $aia\_prep.pro$, $hmi\_prep.pro$ and $secchi\_prep.pro$ function, which process the images by calibrating, rotating, and centering.
We downsample them to be same resolution ($1024 \times 1024$ pixels), and the solar radius ($R_{\odot}$) is fixed at 450 pixels.
We mask the area outside $0.98 R_{\odot}$ of disk center to minimize the uncertainty of limb data.
For the calibration of the EUV data, all data numbers are scaled by median values of the original data on the solar disk, which are fixed at 100.
Then the logarithms of the scaled data are normalized from -1 to 1 with the saturation values of $0$ (lower limit) and $log(200)$ (upper limit).
Then we combine the three passband images from SDO and the STEREOs into the RGB channel dimensions.
The magnetograms for training have an upper and lower saturation limit of $\pm 3000$ Gauss for the normalization.
Finally, we manually exclude a set of SDO data pairs and STEREO data with poor quality; for example, noise images because of solar flares, those with incorrect header information, those with infrequent events such as eclipses, transits, etc..

We use HMI daily updated radial field synoptic map with polar field correction \citep{Sun2011} for the conventional magnetic field map at the photosphere.
Hereafter, this is called the HMI synoptic map, and is provided by the Joint Science Operation Center (JSOC).
The synoptic map well represents the region within $\pm60 ^{ \circ}$ of longitude with a daily updated frontside magnetogram.
However, the map still has several uncertainties at the farside.
Here, we improve the farside of the synoptic map in Section \ref{sec:synchronic}.

%%%%%%%%%%%%%%%%%%%%%%%%%%%%%%%%%%%%%%%%%%%%%%%%%%%%%%%%%%%%%%%%%%%%%%%%%%%%%%%%

\section{Method} \label{sec:Method}

\subsection{Pix2PixHD model} \label{sec:pix2pixHD}

%%%%%%%%%%%%%%%%%%%%%%%%%%%%%%%%%%%%%%%%%%%%%%%%%%%%%%%%%%%%%%%%%%%%%%%%%%%%%%%%
\begin{figure*}[t]
\includegraphics[scale=0.57]{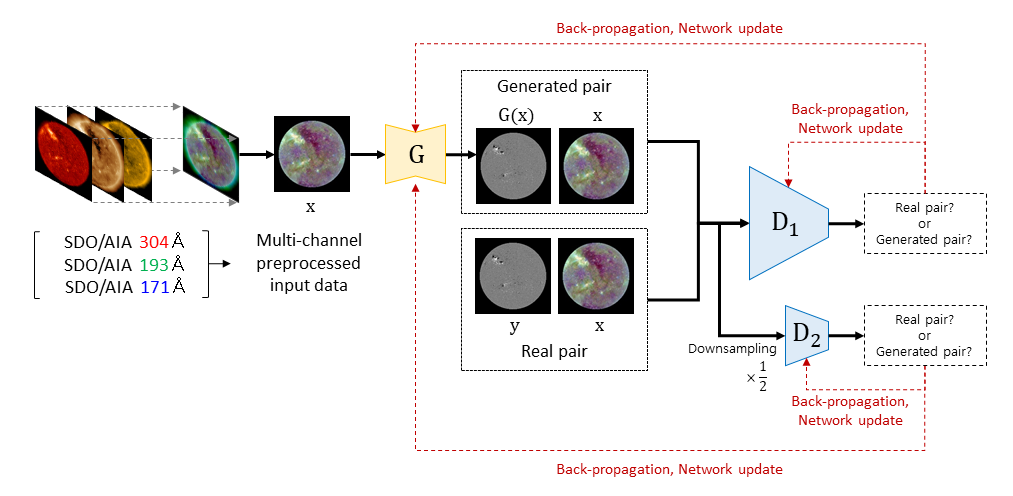}
\centering
\caption{
Flowchart and structures of our deep-learning model.
$G$ is the generator, $D_{1}$ and $D_{2}$ are the discriminators, $x$ is an EUV image with three passbands, $y$ is a real magnetogram, and $G(x)$ is an AI-generated magnetogram. 
}
\label{figure01}
\end{figure*}
%%%%%%%%%%%%%%%%%%%%%%%%%%%%%%%%%%%%%%%%%%%%%%%%%%%%%%%%%%%%%%%%%%%%%%%%%%%%%%%%

We adopt the \textit{Pix2PixHD} model, which is one of the popular deep-learning methods for image translation of high-resolution images without significant artifacts, to generate solar farside magnetograms.
The \textit{Pix2PixHD} consists of two major networks: one is a generative network (generator) and the other is a discriminative network (discriminator).
The generator tries to generate realistic output from input, and the discriminator tries to distinguish the more realistic pair between a real pair and a fake pair.
The real pair consists of a real input and a real output.
The fake pair consists of a real input and an output from the generator.
We construct a real input, a real output, and an output from the generator as a combined image of three EUV passbands, a magnetogram, and an AI-generated magnetogram, respectively.

The generative network consists of several convolutional layers and transposed convolutional layers.
A convolutional layer contains a set of filters that extract features automatically from the input data, like a human visual system, and whose parameters are learned or updated during the model training process.
A transposed convolutional layer is an inverse process of convolution and tries to reconstruct output from the extracted features.
The discriminative network is a classifier that consists of several convolution layers.
Features which are passed through the convolution layers are fed into a single sigmoid output in order to produce a probability output in the range of 0 (fake) to 1 (real) so that the discriminator acts like a classifier.

While the model is training, both networks compete with each other and get an update at every step with loss functions.
Loss functions are objectives that score the quality of results by the model, and the network automatically learns that they are appropriate for satisfying a goal, i.e., the generation of realistic magnetograms.
We train and evaluate our model using SDO frontside data pairs, and generate farside magnetograms from STEREO's EUV observations as an input of the generator.

Networks of the \textit{Pix2PixHD} model get an update with two loss functions: one is a conditional generative adversarial network (cGAN) loss ($\mathcal{L}_{cGAN}$), and the other is a feature-matching loss ($\mathcal{L}_{FM}$).
The cGAN loss is a basic function of the models based on cGANs, and aim for the generator and discriminator to compete.
In order to clearly discriminate the real and fake pairs, the discriminator tries to maximize the loss.
The generator tries to generate realistic data that fools the discriminator, thus minimizing the loss.
The $\mathcal{L}_{cGAN}$ gets a lower value when $D(x,G(x))$ has a value close to 1, in which the AI-generated image is similar to the real output.
The loss of cGAN is given by
\begin{equation}\label{LcGAN}
\mathcal{L}_{cGAN}(G,D) = log\Big(D(x,y)\Big) + log\Big(1-D(x,G(x))\Big),
\end{equation}
where $x$, $y$, and $G(x)$ are a real input, a real output, and an output from the generator, respectively.
The $\mathcal{L}_{FM}$ is to regularize the fake pair to have more similar distribution and statistics with the real pair from multiple layers of the discriminator.
The loss of feature matching is given by 
\begin{equation}\label{Lfm}
\mathcal{L}_{FM}(G,D) = \sum_{i=1}^{T} \frac{1}{N_{i}} \Bigg[ \Big\|D^{(i)} (x,y)-D^{(i)} (x,G(x)) \Big\| \Bigg],
\end{equation}
where $T$, $i$, and $N_{i}$ are the total number of layers, a serial number of the layers, and the number of pixels in output feature maps of each layer, respectively.

Figure \ref{figure01} shows the main structure of our model, which has one generator and two discriminators ($D_{1}$ and $D_{2}$).
One discriminator gets input pairs of the original pixel size, and the other gets input pairs that are downsampled by half.
Each discriminator classifies real pairs and generated pairs with different scales, and guides the generative networks to generate globally consistent images and produce finer details.  
Their full loss function combines both cGAN loss and feature-matching loss.
We use 10 for a relative weight ($\lambda$) which controls the importance of $\mathcal{L}_{cGAN}$ and $\mathcal{L}_{FM}$ as in \cite{Wang2018}.
When the model is trained, the generator tries to minimize the full loss and the discriminators try to maximize the cGAN loss (equation \ref{Lfinal}).
To minimize or maximize the loss, we use Adam solver \citep{Kingma2015} as an optimizer for both the generator and the discriminator.

\begin{equation}\label{Lfinal}
\min_{G} \Bigg(\Big(\max_{D_{1},D_{2}} \sum_{k=1,2} \mathcal{L}_{cGAN} (G, D_{k})\Big) + \lambda \sum_{k=1,2} \mathcal{L}_{FM} (G, D_{k})\Bigg)
\end{equation}

In our model, an input image ($x$) of three EUV channels is given to a generator and it generates an HMI-like magnetogram ($G(x)$).
The model calculates $\mathcal{L}_{cGAN}$ and $\mathcal{L}_{FM}$ from the results of discriminators.
Then the networks get an update at every step with the losses and they are iterated until the assigned iteration, which is a sufficient number assuring the convergence of the model.
We train the model with 3412 pairs of training dataset for 630,000 iterations ($\sim150$ epochs), and save AI-generated magnetograms from the evaluation inputs every 10,000 iterations.
Our code is available at \url{https://github.com/JeongHyunJin/Jeong2020_SolarFarsideMagnetograms}.
In the readme file, we explain the architecture and selected hyperparameters of our deep-learning model.
For basic and extensive information on the deep learning, please refer to \cite{Buduma2017}, \cite{Goodfellow2016}, and \cite{Subramanian2018}.

\subsection{PFSS model} \label{sec:PFSS}

We use PFSS extrapolation, which is a well-established method for estimating large scale structure of global corona \citep{Riley2006}
And the PFSS is much more widely used for the space weather forecast \citep{Hakamada2005, Pomoell2018} than the Nonlinear Force-Free Field (NLFFF) extrapolation and the magnetohydrodynamic (MHD) approach.
The extrapolation calculates current-free field of the corona from the bottom boundary radius to the source surface radius ($R_{SS}$) by solving the Laplace equation, which is given by 

\begin{equation}\label{Bpfss}
\nabla^{2} \Phi_{B} (r) = 0,
\end{equation}
where $\Phi_{B}$ is a scalar potential.
At the source surface, the magnetic field is assumed to be radial, because the outflowing solar wind drags the field out into the heliosphere.
Open field lines arriving at the source surface are associated with coronal holes (CHs) \citep{Lowder2014}.
The CHs are regions of low-intensity emission in EUV and X-rays due to their low density and temperature compared to the surrounding quiet corona.
The solar winds and interplanetary magnetic fields are known to originate from these regions \citep{Wang1996}.
The heliocentric height of $R_{SS}$  has been conventionally assumed to be 2.5 solar radius ($R_{\odot}$). However, the lower $R_{SS}$ produces better results near solar minimum \citep{Lee2011}.
Lowering the $R_{SS}$ in the PFSS model results in more open fluxes and more coronal hole areas.
In this Letter we use two values: $2.0 R_{\odot}$ and $2.5 R_{\odot}$.
The input boundary condition is the measured radial magnetic ﬁelds in the photosphere-like HMI synoptic maps.
We compute the PFSS model on a uniform grid of $155 \times 240 \times 480$ $(r \times \theta \times \phi)$.

%%%%%%%%%%%%%%%%%%%%%%%%%%%%%%%%%%%%%%%%%%%%%%%%%%%%%%%%%%%%%%%%%%%%%%%%%%%%%%%%

\section{Results} \label{sec:Results}

\subsection{Generation of solar farside magnetograms} \label{sec:farside}

%%%%%%%%%%%%%%%%%%%%%%%%%%%%%%%%%%%%%%%%%%%%%%%%%%%%%%%%%%%%%%%%%%%%%%%%%%%%%%%%
\begin{deluxetable*}{ccccccc}[htb!]

%\begin{tabular*}{ccccccc}[htb!]
\tablenum{1}

\tablecaption{Three Objective Measures of Comparison between SDO/HMI Magnetograms and AI-generated Ones for Full Disk, ARs and QRs.\label{tab:Table01}}
\tablewidth{0pt}
\tablehead{
\colhead{} & \multicolumn2c{Full Disk} & \multicolumn2c{AR} & \multicolumn2c{QR}  \\
\cline{2-7}
\colhead{} & \multicolumn2c{825 images} & \multicolumn2c{1,033 patches} & \multicolumn2c{825 patches}  \\
\colhead{} & \multicolumn2c{($1,024\times 1,024$ pixels)} & \multicolumn2c{($128\times 128$ pixels)} & \multicolumn2c{($128\times 128$ pixels)}  \\
\cline{2-7}
\colhead{}  &  \colhead{Ours} & \colhead{KPL19}
&  \colhead{Ours} & \colhead{KPL19} &  \colhead{Ours} & \colhead{KPL19}
}
\startdata
Total unsigned magnetic flux CC & 0.99 & 0.97  & 0.95  & 0.95 & 0.98 & 0.74 \\
Net magnetic flux CC & 0.86   & -  & 0.93 & -  & 0.97  & -  \\
Mean pixel-to-pixel CC ($8 \times 8$ binning) & 0.81  & 0.77  & 0.79  & 0.66 & 0.62 & 0.21  \\
\enddata
\tablecomments{For comparison with the previous research, the results of KPL19 are shown together.}
\end{deluxetable*}
%\end{tabular*}

%%%%%%%%%%%%%%%%%%%%%%%%%%%%%%%%%%%%%%%%%%%%%%%%%%%%%%%%%%%%%%%%%%%%%%%%%%%%%%%%

We train and evaluate our deep-learning model using pairs of SDO/AIA three EUV passband images and SDO/HMI LOS magnetograms.
Table \ref{tab:Table01} shows results of three objective measures for full disk, active regions (ARs) and quiet regions (QRs) between real and AI-generated magnetograms for the evaluation dataset.
First, we calculate correlation coefficients (CCs) between the total unsigned magnetic flux (TUMF) of the SDO/HMI magnetograms with a full dynamic range and that of AI-generated ones.
Our model shows that the CCs are 0.99, 0.95, and 0.98 for 819 full disk, 1,281 ARs and 819 QRs.
These values demonstrate that our model can successfully generate TUMF over all regions.
Second, we calculate CCs between the net magnetic flux (NMF) of SDO/HMI magnetograms and that of AI-generated ones, and those are 0.86, 0.93, and 0.97 for the same data sets, respectively.
Because discrete magnetic field polarity at the limb of solar disk, NMF CC for full disk is lower than those for ARs and QRs.
Third, mean pixel-to-pixel CCs between SDO/HMI magnetograms and that of AI-generated ones after $8 \times 8$ binning are 0.81, 0.79, and 0.62.
These imply that our model greatly improves the generation of magnetograms for not only ARs but also QRs when compared with results of KPL19.
In particular, it is noticeable that the mean pixel-to-pixel CCs of the QRs has greatly increased.

%%%%%%%%%%%%%%%%%%%%%%%%%%%%%%%%%%%%%%%%%%%%%%%%%%%%%%%%%%%%%%%%%%%%%%%%%%%%%%%%
\begin{figure*}[t]
\vspace{15mm}
\includegraphics[scale=0.33]{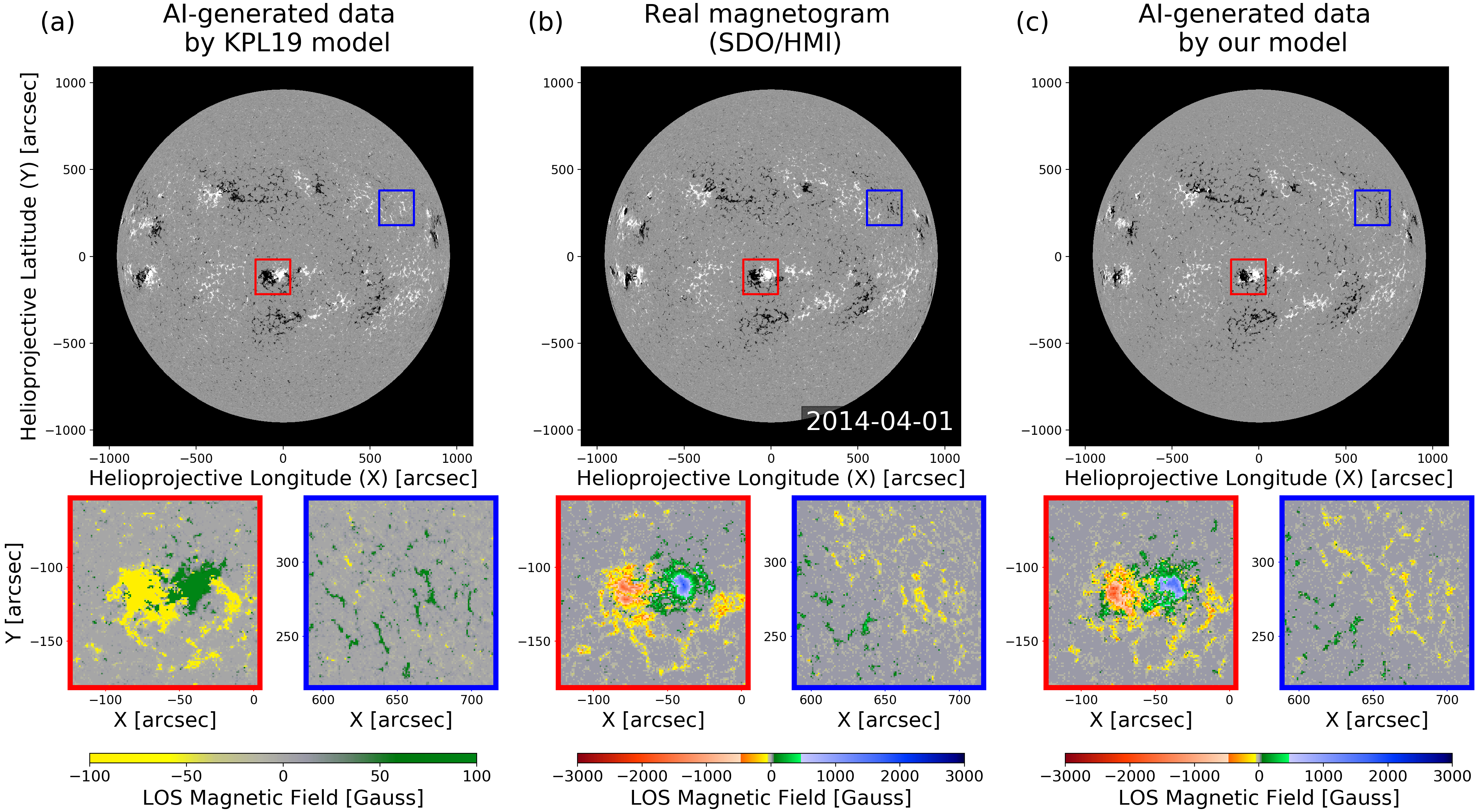}
\centering
\caption{
Comparisons of a real SDO/HMI line-of-sight magnetogram and AI-generated magnetograms.
(a) AI-generated data by KPL19 model. 
(b) real magnetogram on 2014 April 1 at 12:00 UT.
(c) AI-generated one by our model.
 Full-disk magnetograms are displayed as white for positive polarity and black for negative one.
 A solar AR on the center of the solar disk (red box) and a QR on the limb (blue box) are zoomed and represented with other color maps showing the large dynamic range values in Gauss.
}
\label{figure02}
\end{figure*}
%%%%%%%%%%%%%%%%%%%%%%%%%%%%%%%%%%%%%%%%%%%%%%%%%%%%%%%%%%%%%%%%%%%%%%%%%%%%%%%%

Figure \ref{figure02} shows a comparison of magnetograms on 2014 April 1: real one, one by KPL19, and ours.
It is taken from the evaluation dataset.
AI-generated magnetograms from KPL19 and our model show overall magnetic field distributions well.
However, in detailed magnetic structures, our AI-generated magnetogram is much more consistent with the real one than one by KPL19 with a couple of strong points.  
First, the NOAA AR 12021 (red box), which shows strong (higher than $1,000$ Gauss) magnetic fields, are well generated by our model.
Second, the networks of magnetic fields (blue box) are well generated by our model, and the distributions of magnetic polarity look like real one.
Based on those results, our model generates more reliable magnetograms than KPL19.

%%%%%%%%%%%%%%%%%%%%%%%%%%%%%%%%%%%%%%%%%%%%%%%%%%%%%%%%%%%%%%%%%%%%%%%%%%%%%%%%
\begin{figure*}[t]
\vspace{15mm}
\includegraphics[scale=0.31]{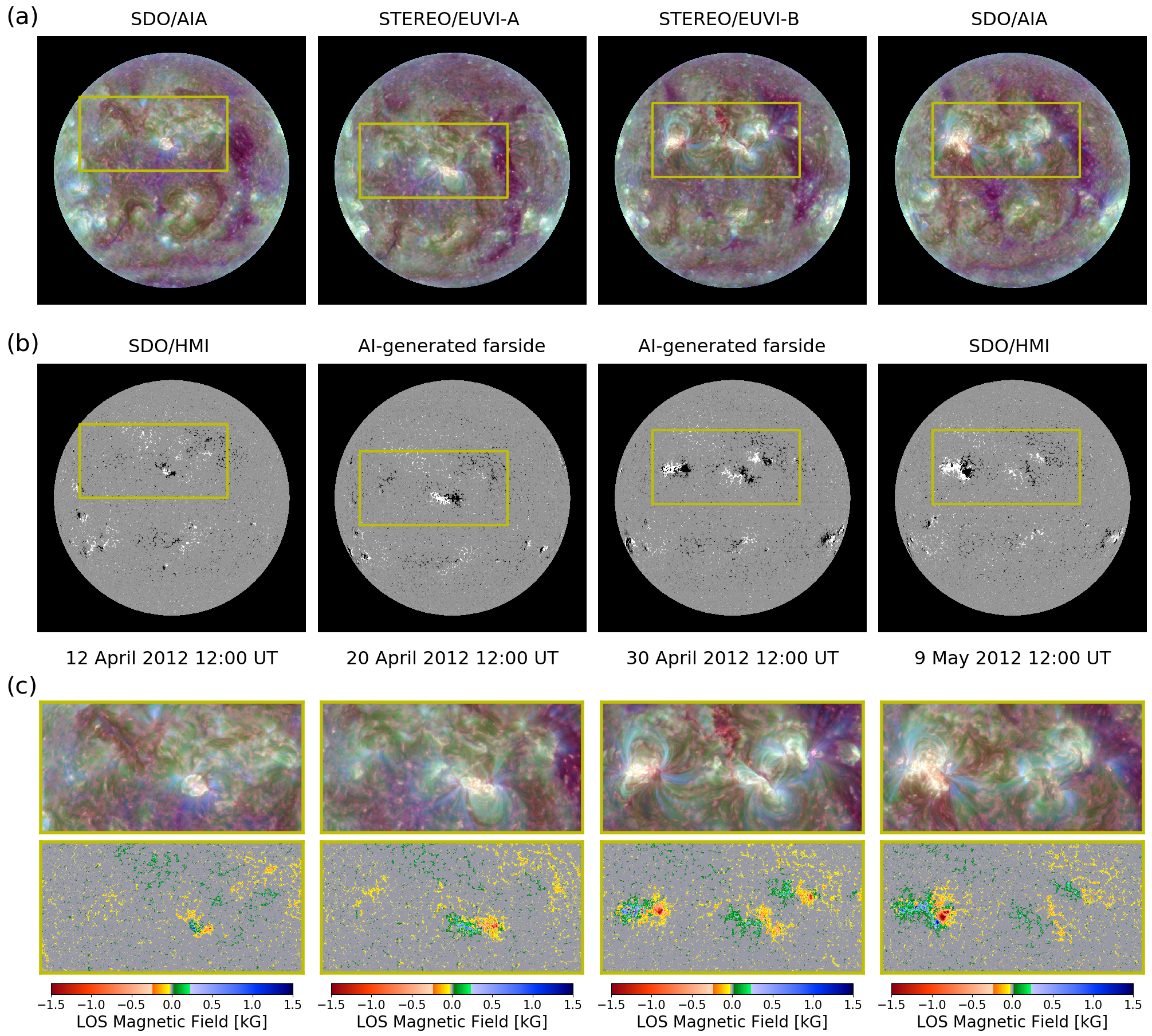}
\centering
\caption{
A series of full-disk EUV images and magnetograms.
(a) The first and fourth EUV images are taken from SDO/AIA 304, 193 and 171 {\AA} passbands.
 The second and third EUV images are taken from STEREO/EUVI A \& B 304, 195 and 171 {\AA} passbands. 
(b) The first and fourth magnetograms are taken from SDO/HMI. 
 The second and third magnetograms are AI-generated farside ones by the model.
 The yellow boxes show the tracking of solar ARs over a solar rotation.
 The positions of the boxes are slightly different due to the inclination angle between the ecliptic plane and orbit of the spacecrafts.
(c) EUV images and magnetograms in the yellow box area are zoomed.
 Full-disk magnetograms are displayed as white for positive polarity and black for negative one.
 The color map of zoomed magnetograms shows large dynamic range values in kG.
}
\label{figure03}
\end{figure*}
%%%%%%%%%%%%%%%%%%%%%%%%%%%%%%%%%%%%%%%%%%%%%%%%%%%%%%%%%%%%%%%%%%%%%%%%%%%%%%%%

We generate more realistic farside magnetograms from the corresponding three EUV passband images of STEREOs by the model.
The images of STEREOs are scaled to SDO ones with a correction factor that is estimated from the ratio of the median on-disk brightness of those images \citep{Ugarte2015,Liewer2017}.
It is not meant as an absolute calibration correction, but it makes consistent conditions with training data for the deep-learning model.
Thus the farside magnetograms generated make it possible to monitor the continuous evolution of solar magnetic field distribution over the solar surface.
Our farside magnetograms generate ARs with realistic magnetic field strength (Figure \ref{figure03}).

%%%%%%%%%%%%%%%%%%%%%%%%%%%%%%%%%%%%%%%%%%%%%%%%%%%%%%%%%%%%%%%%%%%%%%%%%%%%%%%%
\begin{figure*}[t]
\includegraphics[scale=0.64]{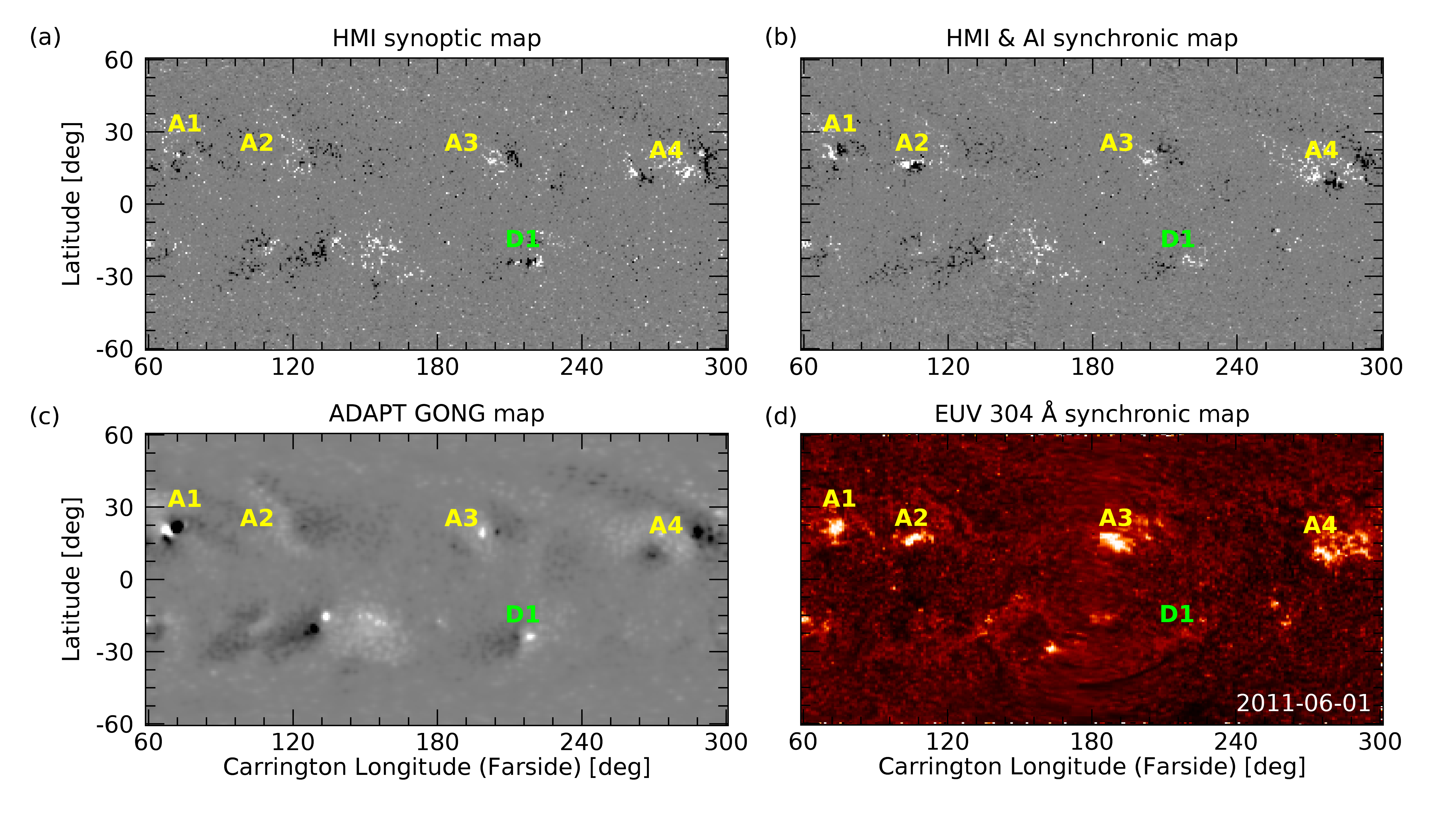}
\centering
\caption{
Comparisons of solar farside magnetic field maps of the photosphere with the EUV 304 {\AA} synchronic map.
(a) Conventional HMI synoptic map on 2011 June 1 at 12:00 UT.
(b) HMI \& AI synchronic map. 
(c) ADAPT GONG map.
(d) EUV 304 {\AA} synchronic map.
  Each map is shows $60 ^{ \circ}$ to $300 ^{ \circ}$ Carrington longitude and within $\pm 60 ^{ \circ}$ latitude.
  "A” indicates appearance or shift case of solar ARs and "D” indicates a disappearance case.
  The polarity of the magnetic fields are shown as a color map of full-disk magnetograms in Figure \ref{figure02} and  \ref{figure03}.
}
\label{figure04}
\end{figure*}
%%%%%%%%%%%%%%%%%%%%%%%%%%%%%%%%%%%%%%%%%%%%%%%%%%%%%%%%%%%%%%%%%%%%%%%%%%%%%%%%

\subsection{Generation of HMI \& AI synchronic maps} \label{sec:synchronic}

We generate the HMI \& AI synchronic map using the farside AI-generated magnetograms.
The AI-generated ones are converted from full disk images to Carrington heliographic coordinated maps, and from the line of sight to the radial magnetic field by applying the radial-acute method \citep{Wang1992} based on their coordinates.
A farside part (from  $60 ^{ \circ}$ to  $300 ^{ \circ}$ Carrington longitude) of the HMI synoptic map is replaced by the AI-generated magnetograms within  $\pm60 ^{ \circ}$ if the farside EUV data are available.
Our synchronic map greatly improves farside magnetic fields, which can be generated at almost the same time as near-real-time EUV observations.

We compare farside photospheric field maps and an EUV 304 {\AA} synchronic map, which is reconstructed with multi-viewpoints observations of SDO/AIA and STEREO/EUVI A and B on 2011 June 1 (Figure \ref{figure04}).
Each map is interpolated to an uniform, $240 \times 480 $ grid in latitude and Carrington longitude for the comparisons of input conditions to compute the coronal field extrapolation. 
The farside of the HMI synoptic map (Figure \ref{figure04}(a))  and the EUV synchronic map (Figure \ref{figure04}(d)) show noticeable differences, because the farside of the synoptic map was taken several days previously.
There were several flux appearance (cases A1, A2, and A3 in Figure \ref{figure04}) and flux disappearance (case D1 in Figure \ref{figure04}) of ARs.
There was also a flux emergence and a shift of the location of an AR moving west (case A4 in Figure \ref{figure04}).
Each location of A1-4 and D1 in Figure \ref{figure04} was set according to the EUV map.
The Air Force Data Assimilative Photospheric flux Transport (ADATP) Global Oscillation Network Group (GONG) map \citep{Hickmann2015} provided by the National Solar Observatory (NSO) is based on a magnetic surface flux transport model, and can predict the changes of sequential magnetic flux that showed at the solar frontside (cases A1 and D1 in Figure \ref{figure04}(c)).
As shown in cases A2 and A3 in Figure \ref{figure04}(c), rapid changes associated with the emergence of new magnetic regions at the limb or farside are not properly reproduced.
Our synchronic map shows well not only the appearance and disappearance of ARs (cases A1-2, A4, and D1 in Figure \ref{figure04}(b)) but also a shift of an AR (case A4 in Figure \ref{figure04}(b)).
In the case of STEREO data unavailability (e.g. case A3 in Figure \ref{figure04}(b)), our result cannot predict the appearance of the AR.

\subsection{Extrapolation of coronal magnetic fields} \label{sec:extrapolation}

%%%%%%%%%%%%%%%%%%%%%%%%%%%%%%%%%%%%%%%%%%%%%%%%%%%%%%%%%%%%%%%%%%%%%%%%%%%%%%%%
\begin{figure*}[t]
\includegraphics[scale=0.38]{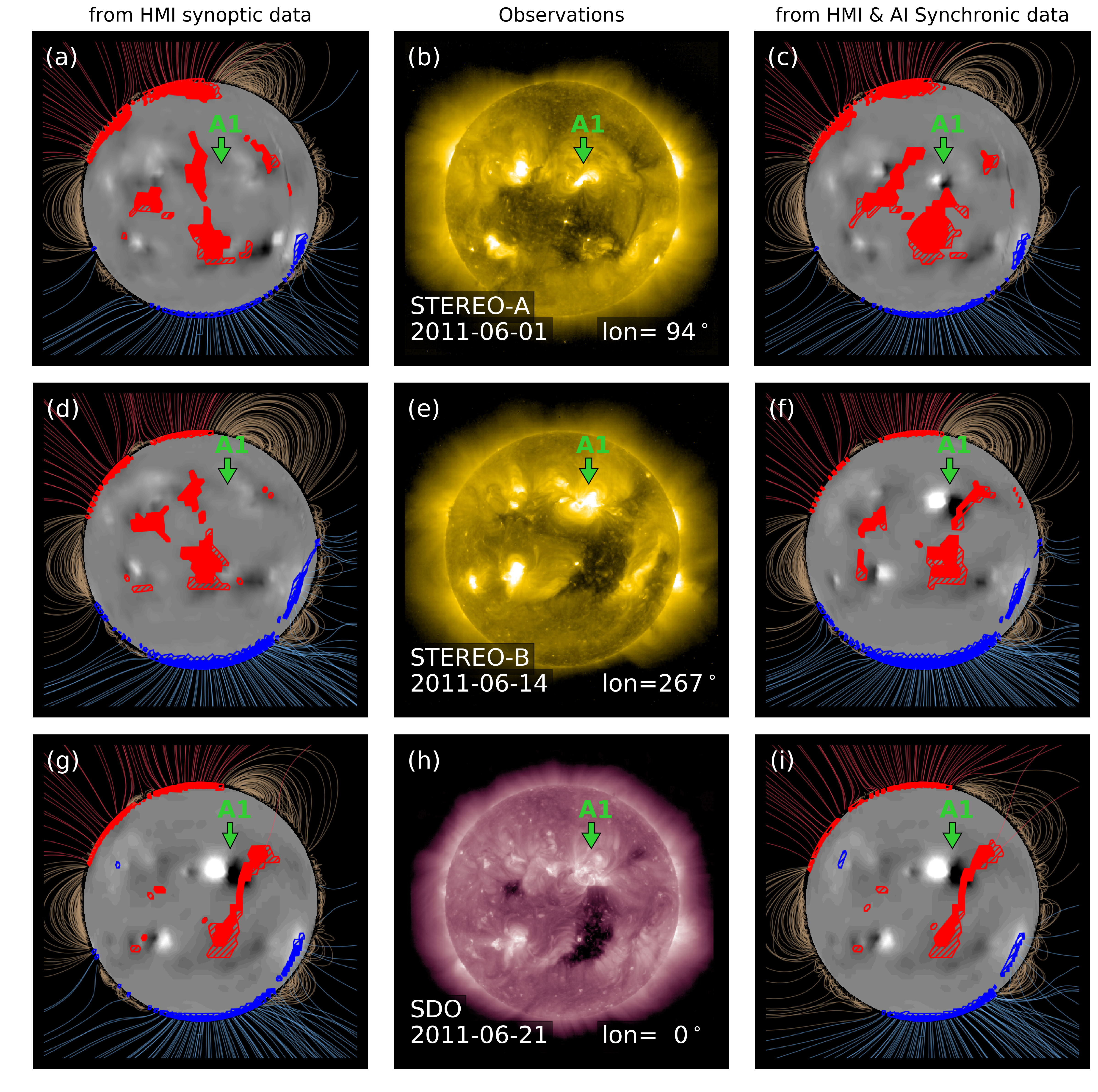}
\centering
\caption{
Comparisons between EUV observations and results of PFSS extrapolations from conventional and HMI \& AI synchronic data in view of full-disk observations.
(a)-(c) Result of PFSS extrapolation from HMI synoptic data, EUV observation, and result of PFSS extrapolation from HMI \& AI synchronic data at the position of STEREO A on 2011 June 1.
(d)-(f) Those at the position of STEREO-B on 2011 June 14.
(g)-(i) Those at the position of SDO on 2011 June 21.
  Positive and negative polarities of the open fields are indicated with blue and red colors.
  Closed field lines are indicated with dark yellow.
  Field lines at the limb and open field area on the surface are only displayed for comparison.
  The PFSS is computed with $R_{SS} = 2.0 R_{\odot}$ and $R_{SS} = 2.5 R_{\odot}$, and those open field areas are displayed with hatched pattern and filled area, respectively.
  The solar surface is filled with the bottom boundary data to show the distribution of ARs.
  The green arrow represents appearances of the NOAA AR 11236.
}
\label{figure05}
\end{figure*}
%%%%%%%%%%%%%%%%%%%%%%%%%%%%%%%%%%%%%%%%%%%%%%%%%%%%%%%%%%%%%%%%%%%%%%%%%%%%%%%%

We use the HMI \& AI synchronic map as a bottom boundary condition to extrapolate global coronal magnetic fields.
Then we predict open field areas (CHs) from the results of extrapolation, and compare those with CHs observed in EUV emissions.
Figure \ref{figure05} shows the results of extrapolations calculated from the HMI synoptic and HMI \& AI synchronic data, and EUV observations of STEREO/EUVI 284 {\AA} and SDO/AIA 211 {\AA} from 2011 June 1 to 21.
Those EUV passbands are not used for training and generation, and ARs and CHs are well identified in those images.
There was a continuous magnetic flux emergence of the NOAA AR 11236 over a solar rotation.
We select the data when the AR was near the center of the solar disk in each spacecraft observation, and indicate them with green arrows.
The first and second rows in Figure \ref{figure05} show STEREO EUV images at the solar farside and the extrapolation results at the corresponding positions.
These positions are $94 ^{ \circ}$ Carrington longitude near the west limb and $267 ^{ \circ}$ Carrington longitude near the east limb of the solar frontside.
There were appearances of two ARs including NOAA AR 11236 near the center of the solar disk observed by STEREO A (Figure \ref{figure05}(b)).
The PFSS extrapolation from HMI synoptic data cannot represent those ARs, and depict long CH structures along the latitudinal direction near the meridian (Figure \ref{figure05}(a)).
On the other hand, the extrapolation from HMI \& AI synchronic data well represents the ARs and CHs under the NOAA AR 11236 (Figure \ref{figure05}(c)).
And 23 days later, flux emergences of the NOAA AR 11236 and another AR on its southeast region were observed by the STEREO-B on 2011 June 14 (Figure \ref{figure05}(e)).
The PFSS extrapolation from HMI synoptic data show similar CH distributions as the extrapolations from several days previous (Figure \ref{figure05}(d)).
The extrapolation from our synchronic data well represents the distributions of the ARs and CHs consistent with the observed images (Figure \ref{figure05}(f)).
The third row in Figure \ref{figure05} shows SDO observations at the frontside and PFSS results at the corresponding position on 2011 June 21.
The results of these calculations are mostly same, because both extrapolations are computed by the observed frontside magnetograms (Figure \ref{figure05}(g) and (i)).
The PFSS results of our synchronic data (Figure \ref{figure05}(c), (f) and (i)) show continuous sequences of ARs and CHs from the farside to the frontside of the Sun.

%%%%%%%%%%%%%%%%%%%%%%%%%%%%%%%%%%%%%%%%%%%%%%%%%%%%%%%%%%%%%%%%%%%%%%%%%%%%%%%%
\begin{figure*}[t]
\includegraphics[scale=0.37]{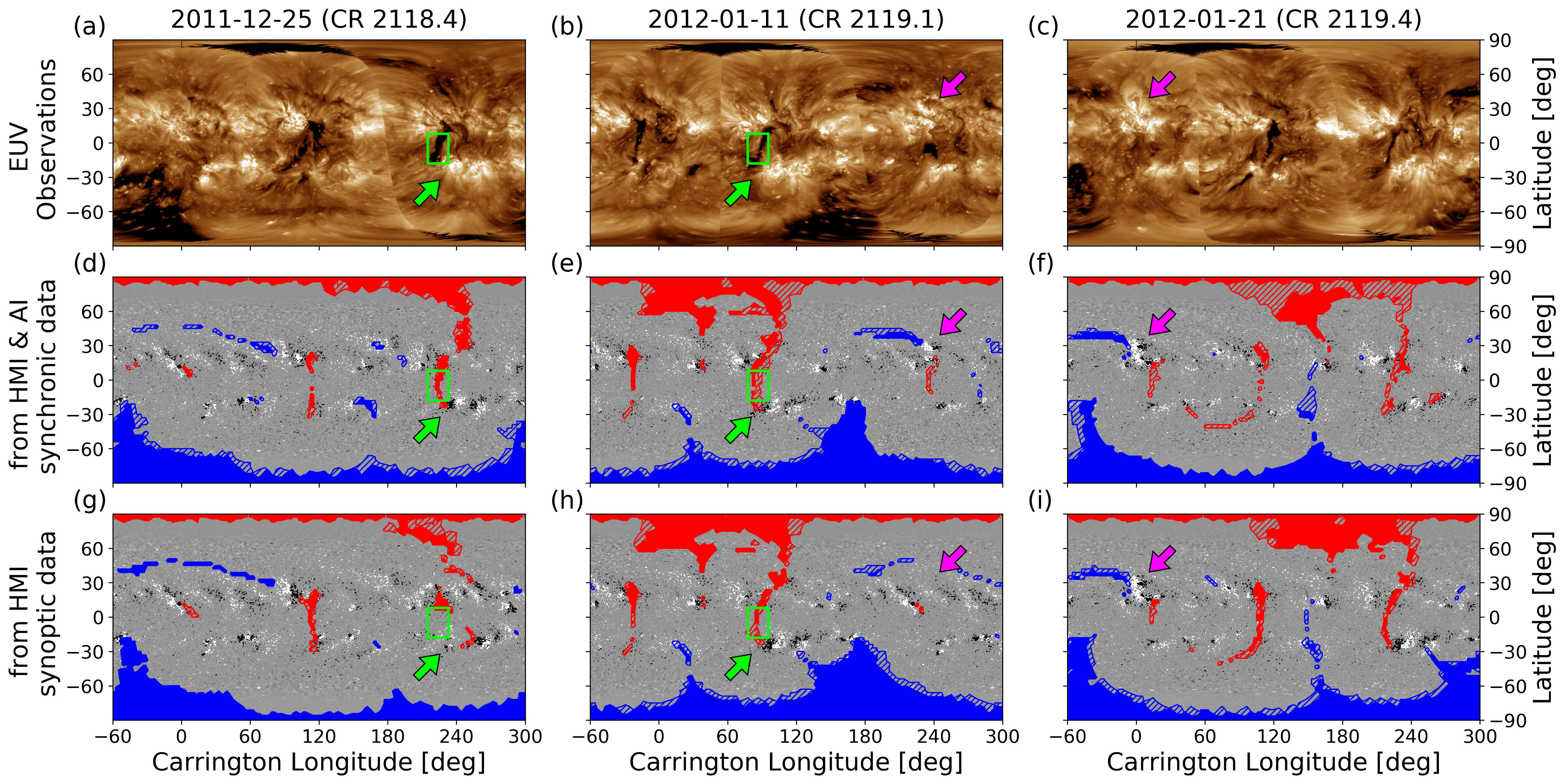}
\centering
\caption{
Comparisons between EUV synchronic maps and results of PFSS extrapolations from conventional and HMI \& AI synchronic data in view of Carrington maps over a solar rotation.
(a)-(c) EUV synchronic maps from STEREO/EUVI A \& B 195 {\AA} and SDO/AIA 193 {\AA} on 2011 December 25, and those on 2012 January 11 and 21.
(d)-(f) Results of PFSS extrapolation from HMI \& AI synchronic data corresponding to the EUV maps.
(g)-(i) Those from HMI synoptic data corresponding to the EUV maps.
  Green and pink arrows indicate two solar ARs linked with equatorial open field regions.
  The boxes with green represent the areas, which include a coronal hole, where we compute the magnetic fluxes.
  Other features are described in Figure \ref{figure05}.
}
\label{figure06}
\end{figure*}
%%%%%%%%%%%%%%%%%%%%%%%%%%%%%%%%%%%%%%%%%%%%%%%%%%%%%%%%%%%%%%%%%%%%%%%%%%%%%%%%

Figure \ref{figure06} shows the EUV observations and CHs identified from the extrapolations in view of Carrington maps from 2011 December 25 to 2012 January 21.
The first row in Figure \ref{figure06} shows synchronized EUV maps from STEREO/EUVI A \& B 195 {\AA} and SDO/AIA 193 {\AA}.
The second row in Figure \ref{figure06} shows the results from our HMI \& AI synchronic data, and the third row in Figure \ref{figure06} corresponds to the results from conventional HMI synoptic data.
There are two major magnetic flux emergence of the ARs.
They are linked with equatorial open field regions (CHs), which are indicated with green and pink arrows in Figure \ref{figure06}.
Those ARs and CHs are more consistent with our results than those of the conventional method.
Our extrapolations show overall consistent distributions of global magnetic field polarities over one solar rotation.

We calculate the magnetic fluxes for each area (green box in Figure \ref{figure06}) including an equatorial coronal hole. 
On 2011 December 25, when the coronal hole is on the farside of the Sun, the total unsigned magnetic flux, positive magnetic flux, and negative magnetic flux from our HMI \& AI synchronic data are $2.1 \times 10^{ 21}$, $6.2 \times 10^{ 20}$, and $1.5 \times 10^{ 21}$ Mx, respectively.
In HMI synoptic data they are $2.3 \times 10^{ 21}$, $1.0 \times 10^{ 21}$, and $1.3 \times 10^{ 21}$ Mx, respectively.
In our case the ratio of the negative magnetic flux to the total unsigned magnetic flux is about 71 \%, while the ratio from the synoptic data is about 55 \%.
For 2012 January 11 data, where the equatorial coronal hole is well predicted, the ratio of our result is about 70 \%, and 71 \% from the synoptic data. 
Our results show more consistent magnetic fluxes of the unipolar region than those from the HMI synoptic data.
These results represent that our AI-generated magnetograms well generate distributions of magnetic fluxes in coronal holes.

%%%%%%%%%%%%%%%%%%%%%%%%%%%%%%%%%%%%%%%%%%%%%%%%%%%%%%%%%%%%%%%%%%%%%%%%%%%%%%%%

\section{Summary and Prospects} \label{sec:summary}

We want to stress on that our new methodology has several positive prospects for the study of solar corona, heliosphere, and space weather.
First, AI-generated farside magnetograms, together with frontside magnetograms, can be used for the long-term evolution of sunspots and solar magnetic fluxes.
In this Letter, we have applied a deep learning model to the generation of solar farside magnetograms in order to extrapolate solar coronal magnetic fields.
We show that the AI-generated magnetograms in our model can generate strong magnetic fields, which show correlations with real magnetograms of full disk, ARs, and QRs that are higher than those of KPL19.
The improvements provide new opportunities for the global magnetic flux studies, such as tracking solar ARs and studying their evolution \citep{Ugarte2015} or studying the time evolution of open and total magnetic fluxes at the solar surface \citep{Solanki2002}.
As deep learning technology advances, our AI-generated data will become more realistic and the applications from the AI-generated ones will show more promising results.

Second, our synchronic map can be better input data for not only the PFSS global field extrapolations but also MHD approaches such as the Magnetohydrodynamic Algorithm outside a Sphere (MAS) model \citep{Mikic2018}.
We show that our maps are more consistent with EUV observations than the conventional photospheric data.
The MAS model has been used to be computed for higher heights of corona than the source surface radius of the PFSS and constructs the physical parameters of corona, which are not only magnetic field vectors but also plasma properties such as mass density, gas pressure, and velocity.
If our results are used for the input of the MAS model, we expect that the model can produce more reasonable solar coronal and heliospheric physical parameters. 

Third, the HMI \& AI synchronic map may be improved with data assimilation methods and photospheric flux transport models, which include the effects of differential rotation, meridional flow, super-granulation, and random background flux.
In our study, we simply replace a farside HMI synoptic map by the AI-generated farside magnetograms.
There are several techniques that assimilate magnetograms into the flux transport model, e.g., the Schrijver and DeRosa model \citep{Schrijver2003} and the ADAPT model \citep{Hickmann2015}.
As shown in their methods, our AI-generated farside magnetograms can be assimilated into the flux transport model, and as a result we expect that the model may show better results than before.
This model has the advantage that the map has a better stable balance in magnetic flux on the boundary of the generated data and on the map close to the boundary.

Fourth, extrapolated coronal magnetic field data can be used for initial boundary conditions of several space weather prediction models.
We have shown that the PFSS extrapolations with the synchronic data are also more consistent with EUV observations than the conventional methods and show continuous sequences of coronal structure changes over several solar rotations.
Our improved PFSS extrapolation data will be useful as better input conditions for the solar wind forecasting \citep{Hakamada2005,Pomoell2018}, which is a major component for space weather.
For example, the Wang-Sheeley-Arge (WSA) solar wind model \citep{Arge2000} has been widely used to forecast the solar wind at 1 AU from the coronal extrapolation.
It provides improved solar wind conditions to heliospheric MHD models, such as an ENLIL model for forecast of corona mass ejection arrivals \citep{Steenburgh2014}.
Our method will provide more accurate solar wind predictions, especially for the farside of the sun and heliosphere.

Fifth, our PFSS extrapolation can be help us study the global environment from the Sun to the interplanetary space.
The extrapolation has been widely used for decades to study interplanetary fields \citep{Schatten1969,Rust2008}, photospheric sources of the solar wind \citep{Wang2003}, and solar energetic particle (SEP) events \citep{Nitta2006,Park2013}.
Recently, the Parker Solar Probe (PSP) \citep{Fox2016}, the first spacecraft to fly into the low solar corona, has been collecting data on magnetic fields, which has been compared with the time series predictions of radial magnetic fields from the PFSS \citep{Bale2019,Panasenco2020}.
When the training data for our deep-learning model have been prepared for this solar cycle, we hope to compare the results with the PSP observations.
Solar Orbiter, which was launched in 2020, is equipped with a wide range of not only in situ but also remote-sensing instruments \citep{Muller2013}, which can be better examined with our method.
Moreover, our study may give us new insight on the global solar research techniques and on how to construct a set of instruments on board future spaceborne solar imaging missions such as the L4 and L5 missions \citep{Vourlidas2015}. 

\acknowledgments
We really appreciate the referee's constructive comments.
We thank the numerous team members who have contributed to the success of the SDO mission, as well as the STEREO mission.
We also thank the Solar Physics Group at Stanford University for their support in providing timely access to HMI data and synoptic maps.
We acknowledge the community efforts devoted to developing the open-source packages that were used in this work. This work was supported by the BK21 plus program through the National Research Foundation (NRF) funded by the Ministry of Education of Korea, the Basic Science Research Program through the NRF funded by the Ministry of Education (NRF-2016R1A2B4013131, NRF-2019R1A2C1002634, NRF-2019R1C1C1004778, NRF-2020R1C1C1003892), the Korea Astronomy and Space Science Institute (KASI) under the R\&D program ‘Study on the Determination of Coronal Physical Quantities using Solar Multi-wavelength Images (project No. 2019-1-850-02)' supervised by the Ministry of Science and ICT, and Institute for Information \& communications Technology Promotion (IITP) grant funded by the Korea government (MSIP; 2018-0-01422, Study on analysis and prediction technique of solar flares).

%%%%%%%%%%%%%%%%%%%%%%%%%%%%%%%%%%%%%%%%%%%%%%%%%%%%%%%%%%%%%%%%%%%%%%%%%%%%%%%%%
\bibliography{main}{}

\begin{thebibliography}{}
\expandafter\ifx\csname natexlab\endcsname\relax\def\natexlab#1{#1}\fi
\providecommand{\url}[1]{\href{#1}{#1}}
\providecommand{\dodoi}[1]{doi:~\href{http://doi.org/#1}{\nolinkurl{#1}}}
\providecommand{\doeprint}[1]{\href{http://ascl.net/#1}{\nolinkurl{http://ascl.net/#1}}}
\providecommand{\doarXiv}[1]{\href{https://arxiv.org/abs/#1}{\nolinkurl{https://arxiv.org/abs/#1}}}

\bibitem[{Amari {et~al.}(2018)Amari, Canou, Aly, Delyon, \&
  Alauzet}]{Amari2018}
Amari, T., Canou, A., Aly, J.-J., Delyon, F., \& Alauzet, F. 2018, Nature, 554,
  211

\bibitem[{Arge \& Pizzo(2000)}]{Arge2000}
Arge, C., \& Pizzo, V. 2000, Journal of Geophysical Research: Space Physics,
  105, 10465

\bibitem[{Bale {et~al.}(2019)Bale, Badman, Bonnell, Bowen, Burgess, Case,
  Cattell, Chandran, Chaston, Chen, {et~al.}}]{Bale2019}
Bale, S., Badman, S., Bonnell, J., {et~al.} 2019, Nature, 576, 237

\bibitem[{Bertello {et~al.}(2014)Bertello, Pevtsov, Petrie, \&
  Keys}]{Bertello2014}
Bertello, L., Pevtsov, A., Petrie, G., \& Keys, D. 2014, Solar Physics, 289,
  2419

\bibitem[{Buduma \& Locascio(2017)}]{Buduma2017}
Buduma, N., \& Locascio, N. 2017, Fundamentals of deep learning: Designing
  next-generation machine intelligence algorithms (" O'Reilly Media, Inc.")

\bibitem[{Cairns {et~al.}(2018)Cairns, Lobzin, Donea, Tingay, McCauley, Oberoi,
  Duffin, Reiner, Hurley-Walker, Kudryavtseva, {et~al.}}]{Cairns2018}
Cairns, I.~H., Lobzin, V., Donea, A., {et~al.} 2018, Scientific reports, 8, 1

\bibitem[{Chen \& Koltun(2017)}]{Chen2017}
Chen, Q., \& Koltun, V. 2017, 2017 IEEE International Conference on Computer
  Vision (ICCV), 1520

\bibitem[{DeVore {et~al.}(1984)DeVore, Sheeley, \& Boris}]{DeVore1984}
DeVore, C.~R., Sheeley, N., \& Boris, J. 1984, Solar physics, 92, 1

\bibitem[{Downs {et~al.}(2012)Downs, Roussev, van~der Holst, Lugaz, \&
  Sokolov}]{Downs2012}
Downs, C., Roussev, I.~I., van~der Holst, B., Lugaz, N., \& Sokolov, I.~V.
  2012, The Astrophysical Journal, 750, 134

\bibitem[{Fox {et~al.}(2016)Fox, Velli, Bale, Decker, Driesman, Howard, Kasper,
  Kinnison, Kusterer, Lario, {et~al.}}]{Fox2016}
Fox, N., Velli, M., Bale, S., {et~al.} 2016, Space Science Reviews, 204, 7

\bibitem[{Freeland \& Handy(1998)}]{Freeland1998}
Freeland, S.~L., \& Handy, B. 1998, Solar Physics, 182, 497

\bibitem[{Goodfellow {et~al.}(2016)Goodfellow, Bengio, Courville, \&
  Bengio}]{Goodfellow2016}
Goodfellow, I., Bengio, Y., Courville, A., \& Bengio, Y. 2016, Deep learning,
  Vol.~1 (MIT press Cambridge)

\bibitem[{Hakamada {et~al.}(2005)Hakamada, Kojima, Ohmi, Tokumaru, \&
  Fujiki}]{Hakamada2005}
Hakamada, K., Kojima, M., Ohmi, T., Tokumaru, M., \& Fujiki, K. 2005, Solar
  Physics, 227, 387

\bibitem[{Hickmann {et~al.}(2015)Hickmann, Godinez, Henney, \&
  Arge}]{Hickmann2015}
Hickmann, K.~S., Godinez, H.~C., Henney, C.~J., \& Arge, C.~N. 2015, Solar
  Physics, 290, 1105

\bibitem[{Howard {et~al.}(2008)Howard, Moses, Vourlidas, Newmark, Socker,
  Plunkett, Korendyke, Cook, Hurley, Davila, {et~al.}}]{Howard2008}
Howard, R.~A., Moses, J., Vourlidas, A., {et~al.} 2008, Space Science Reviews,
  136, 67

\bibitem[{Huang {et~al.}(2018)Huang, Wang, Xu, Liu, Li, \& Dai}]{Huang2018}
Huang, X., Wang, H., Xu, L., {et~al.} 2018, The Astrophysical Journal, 856, 7

\bibitem[{Illarionov \& Tlatov(2018)}]{Illarionov2018}
Illarionov, E.~A., \& Tlatov, A.~G. 2018, Monthly Notices of the Royal
  Astronomical Society, 481, 5014

\bibitem[{Inoue {et~al.}(2018)Inoue, Kusano, B{\"u}chner, \&
  Sk{\'a}la}]{Inoue2018}
Inoue, S., Kusano, K., B{\"u}chner, J., \& Sk{\'a}la, J. 2018, Nature
  communications, 9, 1

\bibitem[{Isola {et~al.}(2017)Isola, Zhu, Zhou, \& Efros}]{Isola2017}
Isola, P., Zhu, J.-Y., Zhou, T., \& Efros, A.~A. 2017, in Proceedings of the
  IEEE conference on computer vision and pattern recognition, 1125--1134

\bibitem[{Jess {et~al.}(2016)Jess, Reznikova, Ryans, Christian, Keys,
  Mathioudakis, Mackay, Prasad, Banerjee, Grant, {et~al.}}]{Jess2016}
Jess, D.~B., Reznikova, V.~E., Ryans, R.~S., {et~al.} 2016, Nature Physics, 12,
  179

\bibitem[{Ji {et~al.}(2020)Ji, Moon, \& Park}]{Ji2020}
Ji, E.-Y., Moon, Y.-J., \& Park, E. 2020, Space Weather, 18, e2019SW002411

\bibitem[{Kaiser {et~al.}(2008)Kaiser, Kucera, Davila, Cyr, Guhathakurta, \&
  Christian}]{Kaiser2008}
Kaiser, M.~L., Kucera, T., Davila, J., {et~al.} 2008, Space Science Reviews,
  136, 5

\bibitem[{Kim {et~al.}(2019)Kim, Park, Lee, Moon, Bae, Lim, Jang, Kim, Cho,
  Choi, {et~al.}}]{Kim2019}
Kim, T., Park, E., Lee, H., {et~al.} 2019, Nature Astronomy, 3, 397

\bibitem[{Kingma \& Ba(2015)}]{Kingma2015}
Kingma, D.~P., \& Ba, J. 2015, CoRR, abs/1412.6980

\bibitem[{Lee {et~al.}(2011)Lee, Luhmann, Hoeksema, Sun, Arge, \&
  de~Pater}]{Lee2011}
Lee, C., Luhmann, J., Hoeksema, J., {et~al.} 2011, Solar Physics, 269, 367

\bibitem[{Lemen {et~al.}(2011)Lemen, Akin, Boerner, Chou, Drake, Duncan,
  Edwards, Friedlaender, Heyman, Hurlburt, {et~al.}}]{Lemen2012}
Lemen, J.~R., Akin, D.~J., Boerner, P.~F., {et~al.} 2011, in The solar dynamics
  observatory (Springer), 17--40

\bibitem[{Liewer {et~al.}(2017)Liewer, Qiu, \& Lindsey}]{Liewer2017}
Liewer, P., Qiu, J., \& Lindsey, C. 2017, Solar Physics, 292, 146

\bibitem[{Lowder {et~al.}(2014)Lowder, Qiu, Leamon, \& Liu}]{Lowder2014}
Lowder, C., Qiu, J., Leamon, R., \& Liu, Y. 2014, The Astrophysical Journal,
  783, 142

\bibitem[{Miki{\'c} {et~al.}(2018)Miki{\'c}, Downs, Linker, Caplan, Mackay,
  Upton, Riley, Lionello, T{\"o}r{\"o}k, Titov, {et~al.}}]{Mikic2018}
Miki{\'c}, Z., Downs, C., Linker, J.~A., {et~al.} 2018, Nature Astronomy, 2,
  913

\bibitem[{Mirza \& Osindero(2014)}]{Mirza2014}
Mirza, M., \& Osindero, S. 2014, CoRR, abs/1411.1784

\bibitem[{Mueller {et~al.}(2013)Mueller, Marsden, Cyr, Gilbert,
  {et~al.}}]{Muller2013}
Mueller, D., Marsden, R.~G., Cyr, O.~S., Gilbert, H.~R., {et~al.} 2013, Solar
  Physics, 285, 25

\bibitem[{Nandy {et~al.}(2018)Nandy, Bhowmik, Yeates, Panda, Tarafder, \&
  Dash}]{Nandy2018}
Nandy, D., Bhowmik, P., Yeates, A.~R., {et~al.} 2018, The Astrophysical
  Journal, 853, 72

\bibitem[{Nitta {et~al.}(2006)Nitta, Reames, DeRosa, Liu, Yashiro, \&
  Gopalswamy}]{Nitta2006}
Nitta, N.~V., Reames, D.~V., DeRosa, M.~L., {et~al.} 2006, The Astrophysical
  Journal, 650, 438

\bibitem[{Panasenco {et~al.}(2020)Panasenco, Velli, D’Amicis, Shi,
  R{\'e}ville, Bale, Badman, Kasper, Korreck, Bonnell,
  {et~al.}}]{Panasenco2020}
Panasenco, O., Velli, M., D’Amicis, R., {et~al.} 2020, The Astrophysical
  Journal Supplement Series, 246, 54

\bibitem[{Park {et~al.}(2019)Park, Moon, Lee, Lee, Lim, Shin, Kim,
  {et~al.}}]{Park2019}
Park, E., Moon, Y.-J., Lee, J.-Y., {et~al.} 2019, The Astrophysical Journal
  Letters, 884, L23

\bibitem[{Park {et~al.}(2020)Park, Moon, Lim, \& Lee}]{Park2020}
Park, E., Moon, Y.-J., Lim, D., \& Lee, H. 2020, The Astrophysical Journal
  Letters, 891, L4

\bibitem[{Park {et~al.}(2018)Park, Moon, Shin, Yi, Lim, Lee, \&
  Shin}]{Park2018}
Park, E., Moon, Y.-J., Shin, S., {et~al.} 2018, The Astrophysical Journal, 869,
  91

\bibitem[{Park {et~al.}(2013)Park, Innes, Bucik, \& Moon}]{Park2013}
Park, J., Innes, D., Bucik, R., \& Moon, Y.-J. 2013, The Astrophysical Journal,
  779, 184

\bibitem[{Pastor~Yabar {et~al.}(2015)Pastor~Yabar, Mart{\'\i}nez~Gonz{\'a}lez,
  \& Collados}]{Pastor2015}
Pastor~Yabar, A., Mart{\'\i}nez~Gonz{\'a}lez, M., \& Collados, M. 2015, Monthly
  Notices of the Royal Astronomical Society: Letters, 453, L69

\bibitem[{Pesnell {et~al.}(2011)Pesnell, Thompson, \& Chamberlin}]{Pesnell2012}
Pesnell, W.~D., Thompson, B.~J., \& Chamberlin, P. 2011, in The Solar Dynamics
  Observatory (Springer), 3--15

\bibitem[{Pomoell \& Poedts(2018)}]{Pomoell2018}
Pomoell, J., \& Poedts, S. 2018, Journal of Space Weather and Space Climate, 8,
  A35

\bibitem[{Riley {et~al.}(2006)Riley, Linker, Miki{\'c}, Lionello, Ledvina, \&
  Luhmann}]{Riley2006}
Riley, P., Linker, J., Miki{\'c}, Z., {et~al.} 2006, The Astrophysical Journal,
  653, 1510

\bibitem[{Rust {et~al.}(2008)Rust, Haggerty, Georgoulis, Sheeley, Wang, DeRosa,
  \& Schrijver}]{Rust2008}
Rust, D.~M., Haggerty, D.~K., Georgoulis, M.~K., {et~al.} 2008, The
  Astrophysical Journal, 687, 635

\bibitem[{Schatten {et~al.}(1969)Schatten, Wilcox, \& Ness}]{Schatten1969}
Schatten, K.~H., Wilcox, J.~M., \& Ness, N.~F. 1969, Solar Physics, 6, 442

\bibitem[{Scherrer {et~al.}(2012)Scherrer, Schou, Bush, Kosovichev, Bogart,
  Hoeksema, Liu, Duvall, Zhao, Schrijver, {et~al.}}]{Schou2012}
Scherrer, P.~H., Schou, J., Bush, R., {et~al.} 2012, Solar Physics, 275, 207

\bibitem[{Schrijver \& DeRosa(2003)}]{Schrijver2003}
Schrijver, C.~J., \& DeRosa, M.~L. 2003, Solar Physics, 212, 165

\bibitem[{Shin {et~al.}(2020)Shin, Moon, Park, Jeong, Lee, \& Bae}]{Shin2020}
Shin, G., Moon, Y.-J., Park, E., {et~al.} 2020, The Astrophysical Journal
  Letters, 895, L16

\bibitem[{Solanki {et~al.}(2002)Solanki, Sch{\"u}ssler, \&
  Fligge}]{Solanki2002}
Solanki, S., Sch{\"u}ssler, M., \& Fligge, M. 2002, Astronomy \& Astrophysics,
  383, 706

\bibitem[{Steenburgh {et~al.}(2013)Steenburgh, Biesecker, \&
  Millward}]{Steenburgh2014}
Steenburgh, R., Biesecker, D., \& Millward, G. 2013, in Solar Origins of Space
  Weather and Space Climate (Springer), 239--254

\bibitem[{Su \& Van~Ballegooijen(2012)}]{Su2012}
Su, Y., \& Van~Ballegooijen, A. 2012, The Astrophysical Journal, 757, 168

\bibitem[{Subramanian(2018)}]{Subramanian2018}
Subramanian, V. 2018, Deep Learning with PyTorch: A practical approach to
  building neural network models using PyTorch (Packt Publishing Ltd)

\bibitem[{Sun {et~al.}(2011)Sun, Liu, Hoeksema, Hayashi, \& Zhao}]{Sun2011}
Sun, X., Liu, Y., Hoeksema, J., Hayashi, K., \& Zhao, X. 2011, Solar Physics,
  270, 9

\bibitem[{Ugarte-Urra {et~al.}(2015)Ugarte-Urra, Upton, Warren, \&
  Hathaway}]{Ugarte2015}
Ugarte-Urra, I., Upton, L., Warren, H.~P., \& Hathaway, D.~H. 2015, The
  Astrophysical Journal, 815, 90

\bibitem[{Vourlidas(2015)}]{Vourlidas2015}
Vourlidas, A. 2015, Space Weather, 13, 197

\bibitem[{Wang {et~al.}(2018)Wang, Liu, Zhu, Tao, Kautz, \&
  Catanzaro}]{Wang2018}
Wang, T.-C., Liu, M.-Y., Zhu, J.-Y., {et~al.} 2018, in Proceedings of the IEEE
  conference on computer vision and pattern recognition, 8798--8807

\bibitem[{Wang {et~al.}(1996)Wang, Hawley, \& Sheeley}]{Wang1996}
Wang, Y.-M., Hawley, S.~H., \& Sheeley, N.~R. 1996, Science, 271, 464

\bibitem[{Wang \& Sheeley~Jr(1992)}]{Wang1992}
Wang, Y.-M., \& Sheeley~Jr, N. 1992, The Astrophysical Journal, 392, 310

\bibitem[{Wang \& Sheeley~Jr(2003)}]{Wang2003}
---. 2003, The Astrophysical Journal, 587, 818

\end{thebibliography}
\bibliographystyle{aasjournal}
%%%%%%%%%%%%%%%%%%%%%%%%%%%%%%%%%%%%%%%%%%%%%%%%%%%%%%%%%%%%%%%%%%%%%%%%%%%%%%%%%

\end{document}